# Atomistic Insights into the Effects of Phosphorous Doping of Graphene Anode in a Lithium Ion Battery

Babita Rani[a,*], Vladimir Bubanja[b,c] and Vijay K. Jindal[d]

[a]Physics Department, Punjabi University, Patiala-147002, India.

[b]Measurement Standards Laboratory of New Zealand, Callaghan Innovation, PO Box 31310, Lower Hutt, 5040, Wellington, New Zealand.

[c]The Dodd-Walls Centre for Photonic and Quantum Technologies, University of Otago, 730 Cumberland Street, Dunedin, 9016, New Zealand.

[d] Department of Physics, Panjab University, Chandigarh-160014, India.

*Email Address: dr.babita@pbi.ac.in

**Abstract:** Inspired by a recent experimental and theoretical study [Yang et al., 2017], wherein protrusions in graphene have been proposed as an effective strategy to enhance the performance of sodium ion batteries, a comprehensive study of the effects of phosphorus doping in graphene on adsorption and diffusion behaviour of Li is carried out by using density functional theory. We find that protrusion introduced by P-dopant in graphene enhances the adsorption of a single Li atom onto the anode due to an additional partial covalent bonding character between Li and carbon atoms of the substrate. However, with increase in concentration of Li atoms, they tend to form clusters which may lead to dendrite growth and hence battery failure. Finite density of states at Fermi level ensures the electronic conductivity of the P-doped graphene before and after the adsorption of a Li atom. No momentous variation in DOS is observed except a small up shift in Fermi level with increased Li concentration. The presence of such protrusions acts as trapping centres for Li and hinders their migration over the substrate, leading to poor cycling performance of anode. This atomic level study will act as a useful guideline for further development of anode materials for novel battery technologies.

## 1. Introduction

The performance of lithium ion batteries (LIBs) is crucially dependent on the properties of electrode material. Besides graphite [1, 2], various other carbonaceous materials have been investigated for their applications as an anode in LIBs [3-9]. Amongst these, graphene has attracted considerable interest owing to its extraordinary structural, mechanical and electrochemical properties [10]. However, use of graphene as an anode of LIBs has an important instability issue of Li clustering causing dendrite formation, which eventually causes the internal shorting of the cell and its thermal runaway [11]. Li cluster formation can be avoided by improving the binding of Li with the substrate [12]. It has been suggested that adsorption of Li on the substrate can be enhanced by chemically modifying the graphene. Various vacancy and edge defects in graphene have been reported to enhance the adsorption and diffusion properties of Li on the substrate [13, 14]. B-doped mono and di-vacancy defects in graphene lead to optimal adsorption and diffusion properties of Li on the substrate and these systems present themselves as the suitable candidates for anode material [18, 15]. In addition to the above, substitutional doping is also an effective way to modify the surface of graphene. Being similar to C in size, B and N are the commonly used dopants. Graphene doped with B atoms improves the binding of Li with the substrate and prevents cluster formation [16]. On the other hand, adsorption of Li on N-doped graphene is weaker which causes the formation of Li clusters [12]. Very recently, Hu et al. [17] revealed the potential of germagraphene (Ge-doped graphene) for an anode of LIBs on the basis of its high Li storage capacities, low diffusion barrier, low intercalation potential and good cycle performance.

An interesting idea that introduction of protrusions in graphene, instead of making holes, would be an effective way to improve the electrochemical performance of sodium ion batteries, was proposed by Yang et al. [18]. In their joint experimental and theoretical study, they came up with the high-end graphene-based anode in which protrusions were created by substitutional doping of graphene with P atoms. Also, Zhang et al. [19] successfully synthesized P-doped graphene and reported its application as an anode material of LIBs. The electrochemical performance of a battery crucially depends upon the interaction of Li with the anode. In order to provide atomic level picture of this interaction in case of P-doping, the following questions need to be answered. What are the structural effects of P-doping of graphene? How does Li interact with P-doped graphene? Is this interaction similar to the case of N-doped graphene [12], as N and P belong to the same group of periodic table? How is diffusion behaviour of Li affected in the presence of P-dopants? In our study, we find that P-doping causes protrusion which enhances the adsorption of a single Li atom. This adsorption behaviour is found to be different from what was observed in our previous study of N-substituted graphene [12]. The stronger adsorption of Li near the P-dopant causes its trapping which hinders the Li migration. It leads to higher diffusion barrier of Li over the substrate and hence poor cycle performance of anode. With increase in concentration of Li atoms, clusters are formed which although enables higher storage of Li but may result in dendrite growth.

## 2. Computational Method

Ab-initio calculations are carried out by using density functional theory based Vienna Ab-initio Simulation Package (VASP) [20-23]. In these calculations, electron-ion interactions are approximated by employing projector augmented wave (PAW) potentials. Generalized gradient approximation (GGA) parameterized as PW91 [24] is used to describe exchange-correlation interactions of electrons. Brillouin zone is sampled using Monkhorst-Pack k-point mesh of 5x5x1 and an energy cut off of 500 eV is used for plane wave basis expansion. These parameters are found to be sufficient to converge the total energy within a few meV. A super cell of 6x6 (containing 72 atoms) is used to perform full geometry and lattice optimization of graphene. The optimized lattice constant is found to be 2.47 Å. The equilibrium C-C bond length and cohesive energy are 1.42 Å and 8.04 eV respectively. These values are in close agreement with previous experimental value and theoretical predictions [25-28]. Periodic boundary conditions are applied in all directions. The separation between the periodic images in perpendicular direction is set to be greater than 12 Å in order to avoid the interaction between them. The ionic positions are relaxed until forces on each atom are below 0.01 eVÅ$^{-1}$. A denser k-point mesh of 21x21x1 and Gaussian smearing of 0.02 eV is used for electronic structure calculations.

Adsorption energy of Li atom on P-doped graphene is calculated as

$$E_{ads} = E_{GraP-Li} - E_{GraP} - E_{Li} \quad (1)$$

Where $E_{GraP-Li}$ is the total energy of Li adsorbed P-doped graphene, $E_{GraP}$ is the total energy of P-doped graphene and $E_{Li}$ is the total energy of Li atom which is calculated with a single atom in cubic cell of length 10 Å only. Negative value of adsorption energy indicates the stable structure.

## 3. Results and Discussions

### 3.1 Formation and Li adsorption of P-Doped Graphene

We systematically study various adsorption configurations of Li atom(s) on P-doped graphene sheet (PG) of 72 atoms, corresponding to P-concentration of ~1.39 at% which is comparable to that synthesized via a low-cost and scalable thermal annealing method using graphite oxide (GO) and triphenylphosphine (TPP) [19]. Before investigating Li adsorption on P-doped graphene, we first examine the structure of P-doped graphene which is obtained by full structural optimization of graphene with one C-atom substituted with P-atom. We find that doping of P-atom in graphene disturbs its geometric structure to a great extent (figure 1). It is because of its size that P atom does not fit into lattice of graphene but protrudes out of the crystal plane of graphene by 1.32 Å. This protrusion can also be accounted for the change in bonding character between doped atom and the substrate. The bond length between sp$^3$ bonded C and P atom is 1.76 Å which is greater than equilibrium bond length (1.42 Å) between sp$^2$ bonded C atoms of graphene and the bond angle C-P-C is found to be 100.95°. The ease of generation of P-doped graphene can be approximated with the formation energy which is calculated by the formula:

$$E_f = E_{Gra-P} - E_{Gra} + E_C - E_P \quad (2)$$

Where $E_{GraP}$ and $E_{Gra}$ are the total energies of P-doped graphene and pristine graphene respectively. $E_C$ is the energy of one carbon atom in graphene and $E_P$ is the energy of one phosphorus atom. E$_f$ of P-doped graphene is obtained as -0.89 eV which is small and negative indicating that P-doped graphene can easily be formed. These results are in good agreement with those reported in the literature [29, 30]. Bader charge analysis shows that P atom loses its ~3.68e charge to the C atoms of the substrate and become positively charged. The bonded C atoms become negatively charged by obtaining ~1.17e on an average. The protrusion created by doping of P-atom breaks the mirror symmetry of graphene structure. Thus, we have now two sides of graphene labelled as side A and side B (figure 1(b)) where adsorption of Li atom can be considered.

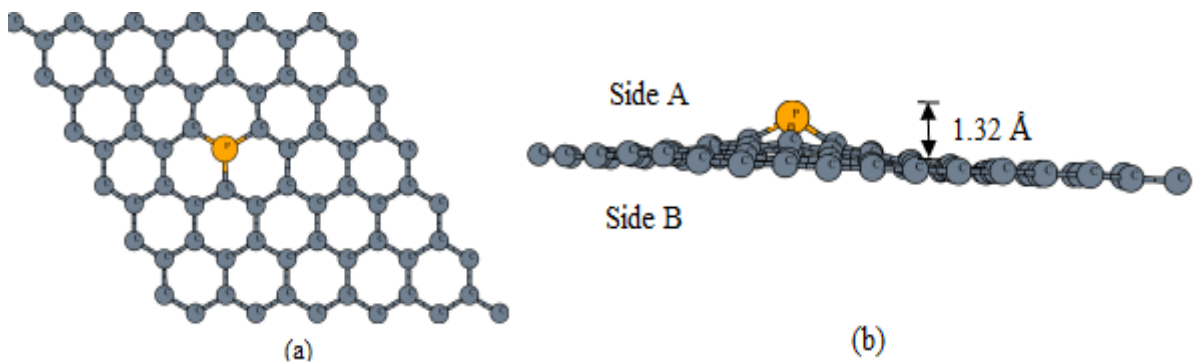

Figure 1 (a) Top and (b) side view of optimized P-doped graphene.

We study different sites for Li to adsorb on the either side of P-doped graphene. The preference for adsorption at a particular site is determined in terms of adsorption energies. We find that on side A of doped graphene, Li prefers to be adsorbed on the hollow site of hexagon near P-atom (figure 2 (a) and (b)) at a distance of 3.17 Å from P-atom with the adsorption energy of -1.83 eV. Bader charge analysis of this adsorption configuration shows that Li atom loses almost its whole valence charge to the bonded C atoms. Further, on side B, Li atom favours to be adsorbed below the P-atom of doped graphene at a distance 2.43 Å with adsorption energy of -2.13 eV (figure 2 (c) and (d)). In this adsorption configuration, Li and P lose ~0.99e and ~3.41e Bader charge respectively to become positively charged, while bonded C atoms, on an average, obtain ~1.28e to become negatively charged. The obtained adsorption energies (in magnitude) are greater than those obtained for pristine graphene (-1.48 eV) and N-doped graphene (-1.00 eV) [12]. This indicates that presence of protrusion caused by P-doping in graphene causes the stronger adsorption of a single Li atom onto the anode. Moreover, Li adsorption energies on either side are greater than the cohesive energy of bulk Li (-1.67 eV) [31] which are indicative of stable adsorption of a Li atom on the substrate. These results are in conformity with the previously reported results [32]. Also, these adsorption configurations are found to be similar to those obtained by Junpin et al. [17] in case of germagraphene (graphene having protrusion created by doping with Ge atom).

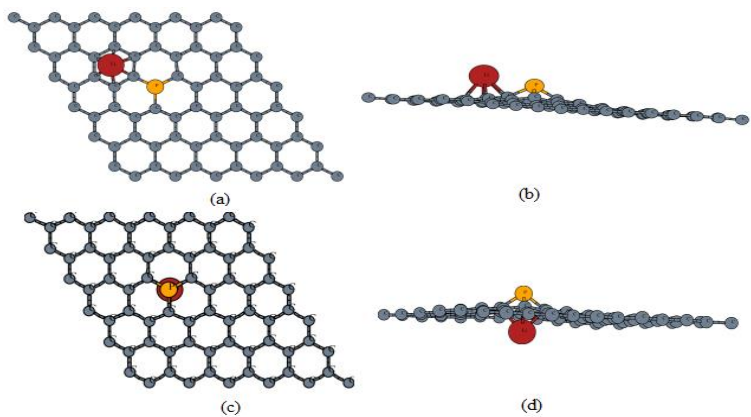

Figure 2 (a) Top and (b) side view of most stable adsorption configuration of Li atom on side A (c) Top and (d) side view of most stable adsorption configuration of Li atom on side B of P-doped graphene.

Li uptake capacity of an anode material is dependent upon the ease of adsorption of Li on the substrate. As a large number of Li atoms are to be stored on the anode, the possibility of Li cluster formation is a primary issue which directly influences the cyclic efficiency of anode. To address this, we systematically increase the number of Li atoms and study the different configurations of Li adsorbed on P-doped graphene in detail. Firstly, we explore all possibilities for the adsorption of two Li atoms near the defect including both atoms on side A, both atoms on side B, one on side A and other on side B; when adsorbed to same hexagon and different hexagons of P-doped graphene. It is found that Li atoms, when initially placed at the centre of two adjacent hexagons containing P on the same side A of P-doped graphene, relax at a distance of 4.59 Å with adsorption energy -1.41 eV/atom (see figure 3(a)). Further, we find that two Li atoms can be adsorbed on side B of the sheet at a distance of 2.83 Å with adsorption energy -1.43 eV/atom (figure 3(b)). Moreover, when Li atoms are initially placed one above and the other below the centre of same hexagon (containing P-atom) of the sheet, they relax in the configuration as shown in the figure 3(c). The adsorption energy in this case is found to be -1.69 eV/atom and Li-Li distance is 4.17 Å. We also find an adsorption configuration (figure 3(d)) with Li-Li distance 5.23 Å when two Li atoms are initially placed at the centre of different adjacent hexagons containing P-atom on the opposite sides of the sheet. The adsorption energy in this configuration is -1.74 eV/atom. It has been found that two side adsorption is energetically favourable in case of two Li atoms. In this adsorption, one Li atom favours to be adsorbed at the centre of a hexagon not containing P atom on side A while on side B, the other stabilizes itself at the hexagon comprising P (figure 3(d)).

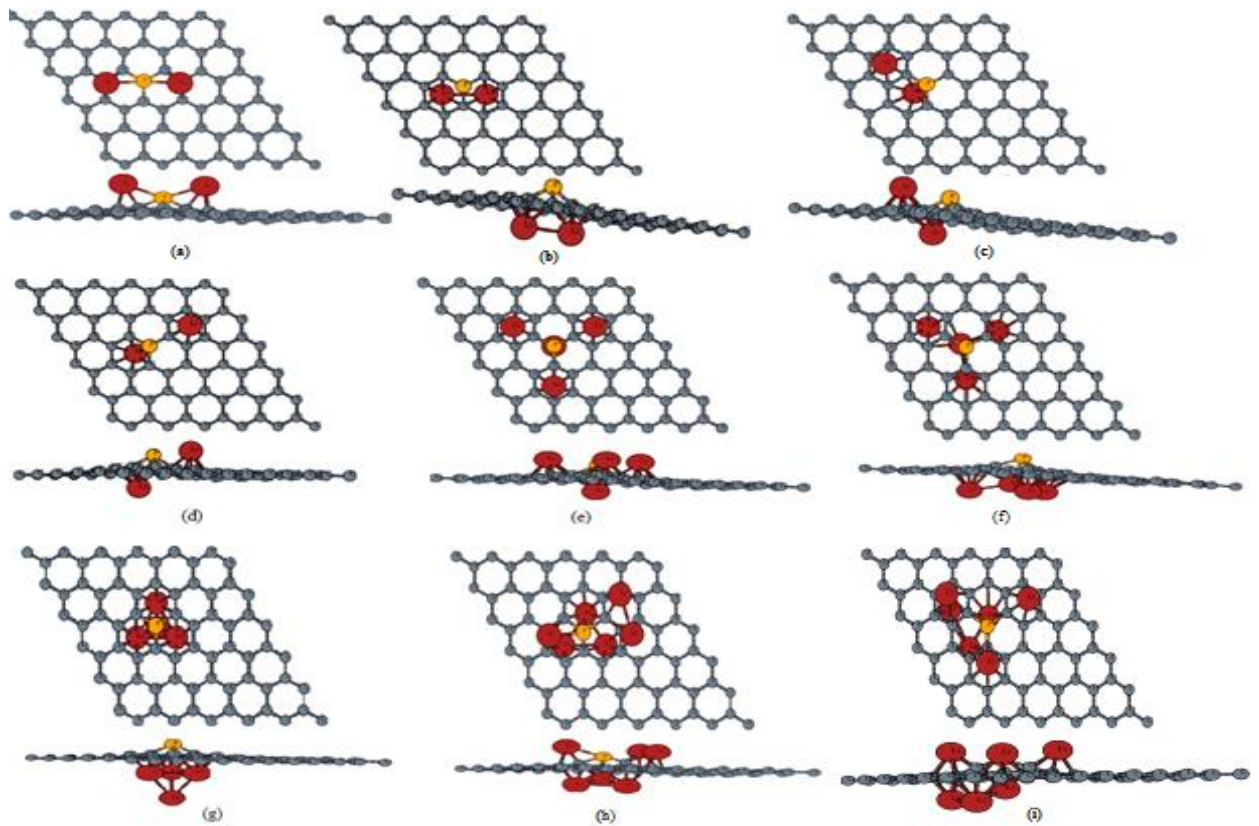

Figure 3 Top and side views of different adsorption configurations of (2, 4 and 6) Li atoms on P-doped graphene.

We further increase the number of Li atoms to four by placing three atoms symmetrically above and one atom below the P-atom of the sheet. The adsorption energy of the two sided adsorption configuration (see figure 3(e)) after relaxation is found to be -1.38 eV/atom with Li-Li distance as 5.40 Å on side A. As distance between Li atoms is greater than 3.04 Å i.e. distance between Li atoms in bulk Li [33], the clustering of Li ions is not likely to occur in this configuration. As the protrusion introduced by P-doping provides extra space for Li to adsorb on side B of the substrate, we also consider two adsorption configurations in which four Li atoms are adsorbed below the defect site as shown in figure 3(f) and (g). Adsorption energies in these configurations are -1.30 eV/atom and -1.52 eV/atom respectively. The corresponding closest Li-Li distances are 2.68 Å and 2.77 Å which are smaller than that of bulk Li i.e. 3.04Å. Moreover, the adsorption energies are smaller than cohesive energy of Li bulk. This indicates the presence of stronger Li intra-cluster bonds relative to the ones in their metallic bulks. The most stable configuration of four atoms is of $Li_4$ cluster adsorbed on side B of P-doped graphene in shown in figure 3(g) in which three Li atoms are in a plane near to the substrate and the fourth one is at an average distance of 3.07 Å from these atoms.

Further, we obtain a configuration (see figure 3(h)) in which six Li atoms are adsorbed. This configuration is obtained by initially placing Li atoms (three below and three above the substrate) at the centre of three adjacent hexagons comprising of P atom. The adsorption energy in this case is -1.27 eV/atom and the closest Li-Li distance is 2.68 Å below the substrate. These parameters again points to the existence of Li clusters as discussed earlier. However, in this case initial Li-Li distance below the substrate may have been too small thereby providing opportune conditions for cluster formation. Therefore, we further investigate the case in which six Li atoms (three below and three above the substrate) are initially placed at comparatively at farther distances. After optimization, we find that three Li atoms again congregate below the protrusion at the closest distance of 2.63 Å (figure 3(i)). Adsorption energy per Li atom of this configuration is -1.31 eV. We didn't go beyond this as the adsorption energies are smaller than the cohesive energy of bulk Li. However, further adsorption of Li ions is possible but it may severely affect the charge/discharge rate due to stability issues posed by Li cluster formation.

From the above, we find that protrusion enhances the adsorption of a single Li atom but as we increase the number of Li atoms, the adsorption energy decreases. Also, Li clusters constitute the most stable state among the different configurations with the same and increased Li concentration. This phenomenon of cluster formation may be accredited to the tendency of Li atoms to adsorb near the extra space created beneath the protrusion. The stability of cluster is portrayed as a result of attractive interaction between Li arising from effective metallic bonding, thereby decreasing the adsorption energy as we increase the number of Li atoms. Maximum uptake capacity of the substrate is found to be four (three on side A and one on side B) near the defect.

**3.2 Charge Density Difference**

Isosurface of charge density difference of different adsorption configurations of single, two and four Li atoms are analysed to elucidate the adsorption. The differential charge density ($\rho_{diff}$) is calculated as

$$\rho_{diff} = \rho_{Li-sub} - \rho_{sub} - \rho_{Li} \qquad (2)$$

where $\rho_{Li\text{-}sub}$, $\rho_{sub}$ and $\rho_{Li}$ are the charge densities of the Li adsorbed P-doped graphene system, P-doped graphene and Li atom(s) respectively.

It is clear from the isosurface plots (figure 4(a) and (b)) of single Li atom adsorbed on the either side of P-doped graphene that charge depletion region is formed around the Li atom which indicates transference of charge from Li to the substrate. Moreover, a detailed examination of the contour plots of charge density difference reveals that charge depletion region is inserted into the accumulation region between Li and C atoms of the substrate in both cases. Therefore, the charge is being shared between Li and C atoms of the substrate which results in an additional and partial covalent bonding character, thereby strengthening the binding of Li with the substrate.

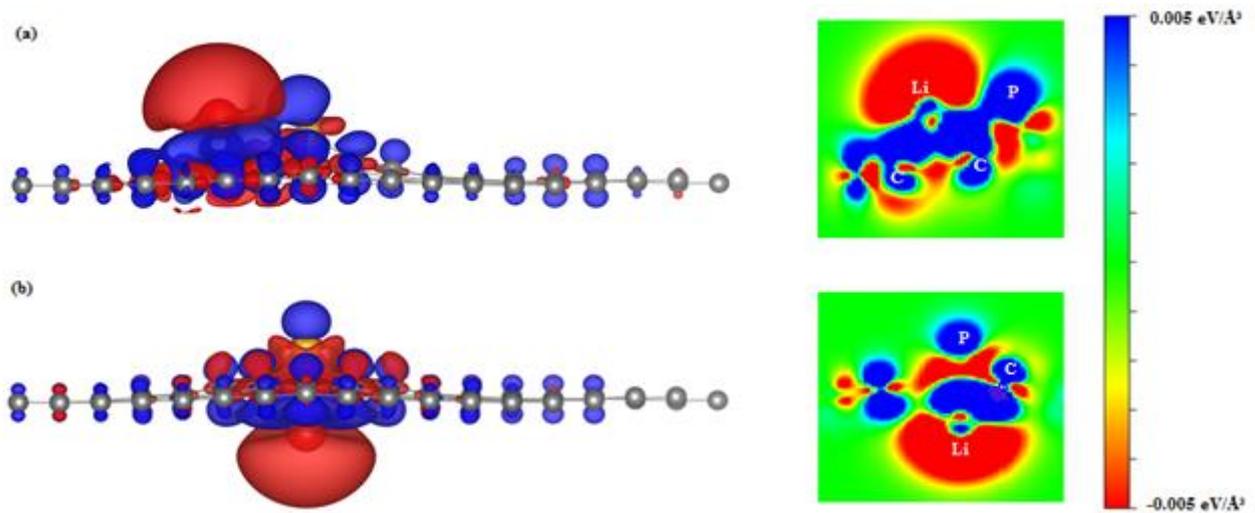

Figure 4 Isosurface and contour of charge density difference of a Li atom adsorbed on (a) side A and (b) side B of P-doped graphene. The plane for the contour is a cross section which comprises of Li, C and doped P atom. The blue and red colours represent charge accumulation and depletion regions respectively. The isovalue is set as 0.005 e/Å³.

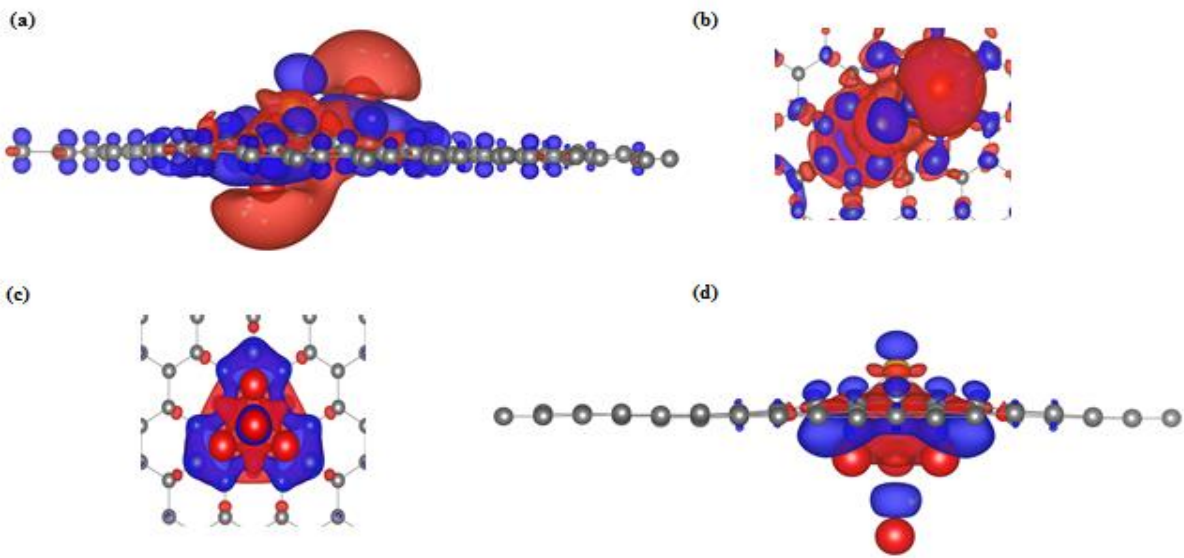

Figure 5 Isosurface of charge density difference of (a, b) two Li atoms (c, d) $Li_4$ cluster adsorbed on P-doped graphene. Blue and red colours represent charge accumulation and depletion regions respectively. The isovalue is set as 0.005 e/Å$^3$

Isosurface plots for the most stable configuration of two atoms are shown in figure 5(a, b). It is clear from these figures that charge is transferred from Li atoms to the substrate. For the most stable adsorption configuration of four Li atoms i.e. $Li_4$ cluster (figure 5 (c, d)), charge transferred from the cluster is distributed at the local region just beneath the protrusion. The charge accumulation region at the centre of cluster may be attributed to the relative large distance between three Li atoms in one plane and the fourth one in another plane.

**3.3 Electronic Conductivity**

The effect of P-doping on the electronic conductivity of graphene anode is determined by calculating the electronic structure of the substrate before and after Li adsorption. Finite DOS at the Fermi level suggests intrinsic conductivity of P-doped graphene (figure 6(a)). These finite DOS are contributed by p-orbitals of phosphorus and carbon atom bonded to it (figure 6(d)). It can be observed from PDOS that hybridization of p-orbitals of carbon and phosphorus atoms occurs below the Fermi level. No significant change in DOS is observed even after the adsorption of a Li atom as seen in figures 6(b) and (c). Also, the substrate remains conductive in both side A and B configurations. This indicates high electron mobility within the material, being not limited by electron transport kinetics [18]. A detailed analysis of PDOS (figure 6(e)) shows that finite DOS at the Fermi level are dominated by contribution from p-orbitals of P with very small and negligible contribution from Li and C orbitals respectively in case of side A adsorption. However, finite DOS in case of side B adsorption are mainly contributed by p-orbitals of C and P atoms including small contribution of Li atom also (see figure 6(f)). Moreover, main peak due to Li orbitals occurs above $E_F$ which corresponds to unoccupied orbitals. The hybridization of C and Li orbital is observed below Fermi energy. These features of DOS confirm the transference of charge from Li to the substrate and presence of an additional partial covalent bonding between Li and the substrate as depicted by charge density difference plots discussed earlier in case of adsorption of a single Li atom on the either side of P-doped graphene (figure 4(a, b)).

As the concentration of Li atoms on P-doped graphene is increased, no effect on DOS is observed except a small shift of Fermi level higher in energies (figure 7(a), (b) and (c)). In case of the most stable two sided adsorption of two Li atoms, combined effects of Li adsorbed on either side of P-doped graphene can be seen (figure 7(d)). In figure 7(e), projected DOS reveal that different orbitals of C, P and Li contibute to finite DOS at Fermi level for two sided adsorption of four Li atoms. For the most stable configuration of four atoms i.e. $Li_4$ cluster adsorbed on side B of P-doped graphene, we find partially filled orbitals of Li atom below Fermi level which is farther from the substrate (figure 7(f)). This apparently describes charge accumulation region in between the Li atoms presented in charge density difference plots (figure 5(a, b)).

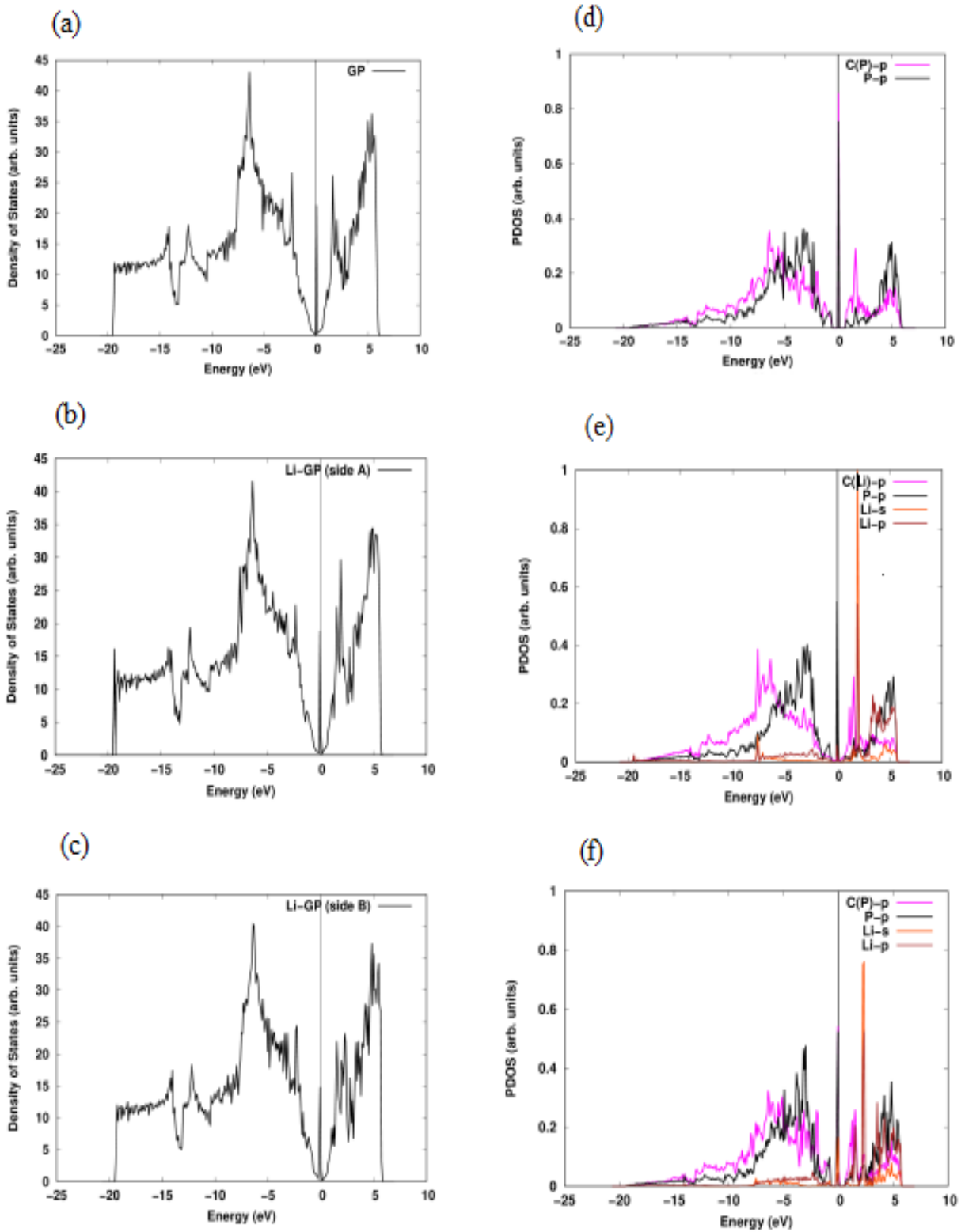

Figure 6 Total density of states of (a) P-doped graphene (b) Li adsorbed P-doped graphene (side A) (c) Li adsorbed P-doped graphene (side B). Projected density of states for (d) P-doped graphene (e) Li adsorbed P-doped graphene (side A) (f) Li adsorbed P-doped graphene (side B). C(P) and C(Li)-p are the p orbital projected DOS of carbon atom bonded to P atom and Li atom respectively. P-p is the p orbital projected DOS of P-atom of P-doped graphene. Li-s and Li-p are s and p orbital projected DOS of Li atom. The energy at Fermi level is set to zero.

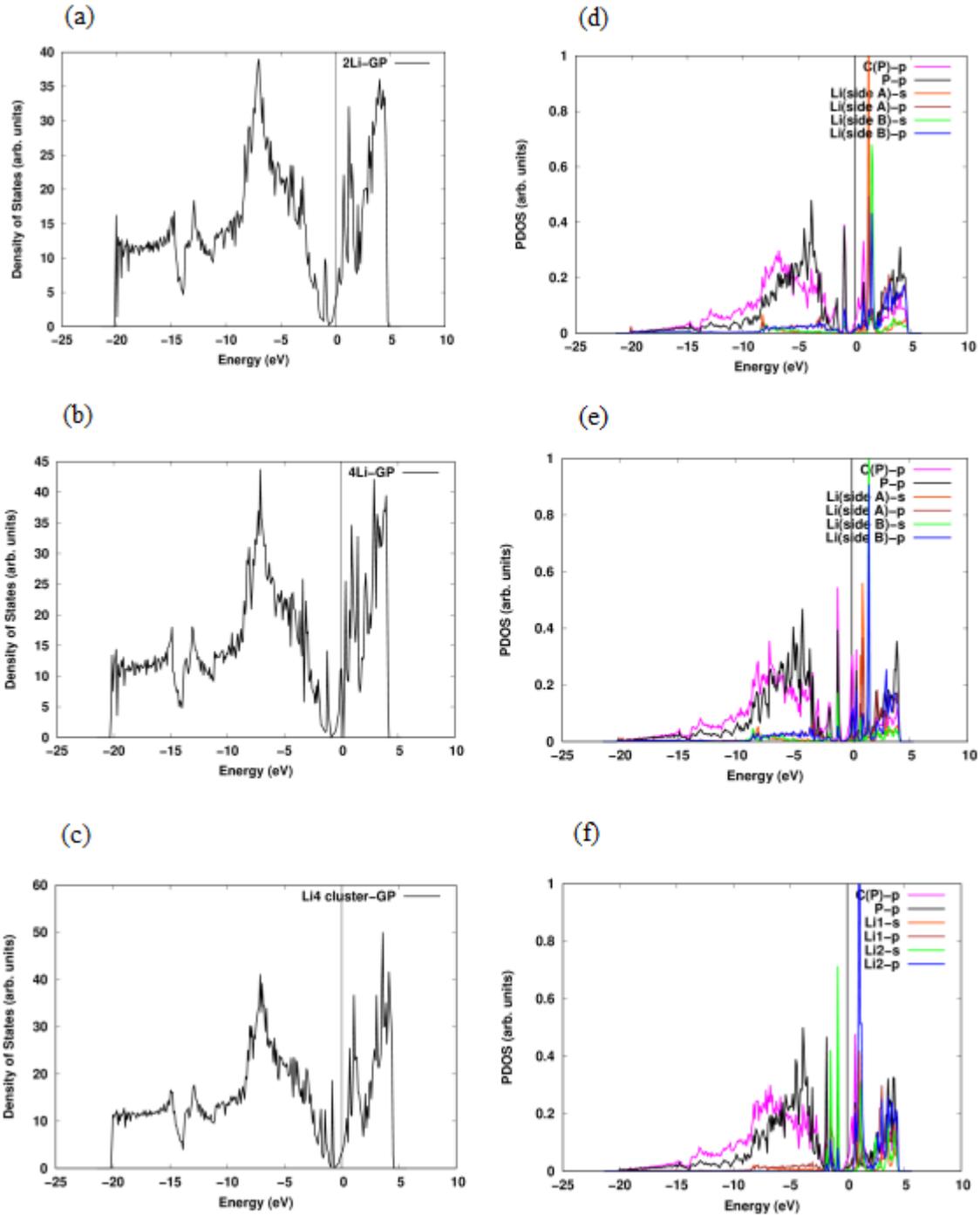

Figure 7 Total density of states of (a) two (b) four Li atoms (c) Li$_4$ cluster adsorbed on P-doped graphene. Projected density of states of (a) two (b) four Li atoms (c) Li$_4$ cluster adsorbed on P-doped graphene. P-p and C(P) are the p orbital projected DOS of P atom and carbon atom bonded to P atom of P-doped graphene. Li(side A)/(side B)-s and Li(side A)/(side B)-p are s and p orbital projected DOS of Li atom adsorbed on side A/side B respectively. Li1-s/p and Li2-s/p are the s/p orbital projected DOS of Li atom of Li$_4$ cluster in the plane nearer and farther from P-doped graphene. The energy at Fermi level is set to zero.

**3.4 Diffusion of Li Atom Across P-doped Graphene**

Efficiency of charge and discharge cycles is determined by diffusion kinetics of Li across the anode material and crucially depends on doping and defects. It was revealed by our previous study [12] that B-doped monovacancy defects in graphene lower the diffusion barrier of Li as compared to those doped with N-atoms. Systematic study of Li adsorption and diffusion on B/N/$V_C$-decorated graphene was performed in [34]. Here, we study the role of extrinsic defects on diffusion kinetics of Li across the substrate. In order to calculate energy barrier for Li to diffuse across the P-doped graphene, we adopt the same formalism as was adopted by Das et al. [15]. The equilibrium configuration of Li on each side of substrate is chosen to be the reference with respect to which the energies of various intermediate configurations along a given diffusion pathway are calculated without any structural relaxation and plotted against reaction coordinates (figure 8) to obtain the energy barrier. These single point calculations are performed to save the computational time. Furthermore, it is imperative to mention here that relaxation does not play any significant role in lowering the barriers [21].

Diffusion of Li on side A is considered along the path indicated in the inset of figure 8(a). It is clear from the figure that starting from its stable adsorption site, Li atom needs to overcome the barrier of height 3.16 eV to diffuse to its equivalent position on another hexagon (i.e. final position along the path shown in the inset of figure 8(a)). Further, three diffusion pathways are chosen across the side B which include all possible non-equivalent paths about the equilibrium position of Li (see the inset of figure 8(b)). These diffusion pathways are referred to as horizontal (path 1) and vertical (path 2 and 3). Diffusion of Li is limited to a distance of 2.46 Å (along path 1) and 2.84 Å (along path 2 and 3) from the equilibrium position of Li. We find that when Li starts diffusing along the horizontal direction (path 1; indicated by blue colour in the inset of figure 8(b)), energy of the system continues to increase and energy difference between the initial and final configurations along this pathway reaches to 3.28 eV (figure 8(b)). Same happens along the path 3 (indicated by green colour in the inset of figure 8(b)) where energy difference reaches to 3.56 eV. Such higher energetic hindrance may be due to trapping of Li at its corresponding adsorption site on the substrate. Further, energy barrier for Li to diffuse along vertical path 2 (indicated by red colour in the inset of figure 8(b)) is found to be 1.39 eV. We observe that the substrate offers an easier path for Li to diffuse along path 2 on side B. However, the calculated migration barriers on either side are much higher than those quoted in the literature for pristine graphene [35]. This suggests that Li is not expected to diffuse readily over the substrate during battery operation, leading to poor cyclic performance of anode.

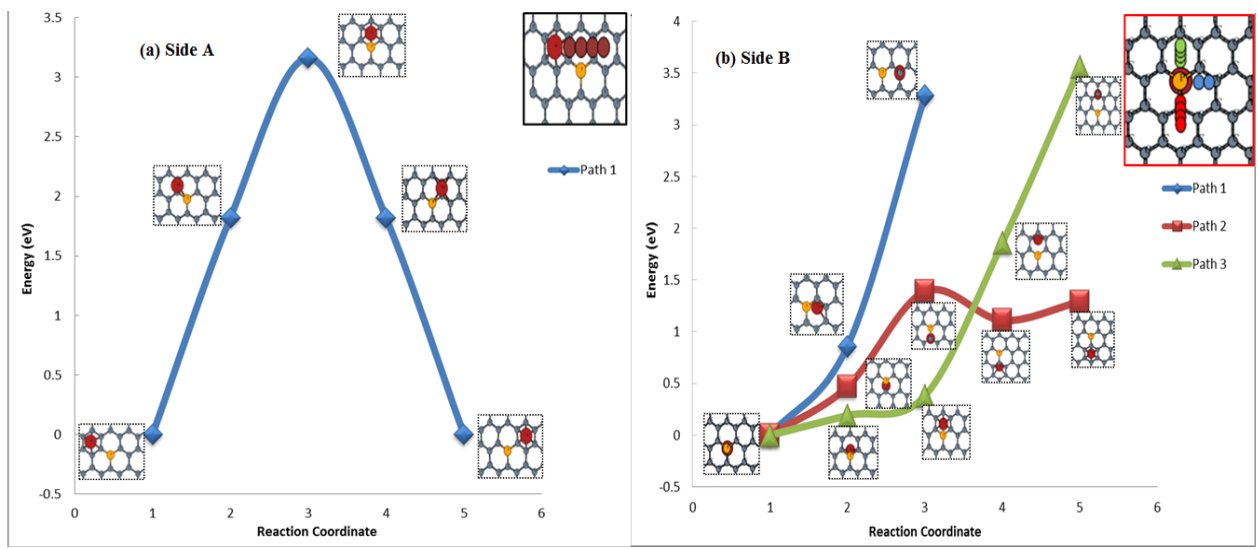

Figure 8 The relative energy of Li at different adsorption sites of P-doped graphene. The insets with dotted border lines represent the corresponding adsorption configuration and solid border lines represent the pathways chosen for Li to diffuse across (a) side A and (b) side B.

## 4. Conclusions

We have carried out an exhaustive study to elucidate the effect of P-doping of graphene anode on the adsorption and diffusion of Li using density functional theory. It is found that P-doping causes protrusion in graphene. The presence of such protrusion causes stronger adsorption of a single Li atom on either side of the substrate owing to additional partial covalent bonding character between Li and carbon atoms as revealed by charge density difference and DOS plots. As observed, although N and P belong to same group of periodic table but their presence as a dopant in graphene reflects different adsorption behaviour of Li over the substrate which may be ascribed to local structural distortion of graphene caused by P-doping. As we increase the concentration of Li atoms, they all favour to be adsorbed in the vicinity of extra space created beneath the protrusion and hence leading to cluster formation. Stability of Li clusters is rendered as a consequence of an efficient metallic bonding, thus decreasing the adsorption energy as we increase the number of Li atoms. Finite density of states at Fermi level ensures the electronic conductivity of the P-doped graphene before and after the adsorption of a single Li atom. No significant change in DOS is observed except a small shift of Fermi level higher in energies with increased concentration of Li atoms. The protrusion acts as trapping centre for Li which obstructs their migration. This leads to higher diffusion barrier of Li over the substrate and hence poor cycling performance of LIB.

**Acknowledgement**

The authors are grateful to the New Zealand eScience Infrastructure (NeSI) for providing the access to the supercomputing facilities.